\documentclass[aps,prb,twocolumn,superscriptaddress,longbibliography]{revtex4-2}
\usepackage{bbm}
\usepackage{amssymb}
\usepackage{graphicx}
\usepackage{amsmath}
\usepackage{natbib}
\usepackage{siunitx}
\usepackage{textcomp}
\usepackage{bm}
\usepackage{color}
\usepackage[hidelinks,colorlinks=true,urlcolor=blue,linkcolor=blue,citecolor=blue]{hyperref}

\def\brho{{\boldsymbol \brho}}
\def\bk{{\boldsymbol k}}

\def\bq{{\boldsymbol q}}
\def\br{\boldsymbol{r}}

\def\la{\langle}

\def\la{\langle}
\def\ra{\rangle}

\begin{document}
\title{Magnetic correlations of a doped and frustrated Hubbard model: benchmarking the two-particle self-consistent theory against a quantum simulator}
\author{Guan-Hua Huang}
\affiliation{Hefei National Laboratory, Hefei 230088, China}
\author{Zhigang Wu}
\email{atticuspku@gmail.com}
\affiliation{Quantum Science Center of Guangdong-Hong Kong-Macao Greater Bay Area (Guangdong), Shenzhen 508045, China}
\date{\today }
\begin{abstract}
Recently a quantum simulator for the 2D Fermi-Hubbard model on an anisotropic triangular lattice has been realized, where both geometrical frustration and doping can be continuously tuned. Here we provide a comprehensive comparison between the magnetic correlations calculated by the two-particle self-consistent (TPSC) theory and those measured in this quantum simulator, at temperatures comparable to the spin exchange energy and Hubbard interactions comparable to the bandwidth. We find overall excellent agreements between the TPSC calculations and the measurements from the quantum simulator at all levels of frustration and doping. This is quite remarkable considering that the Hubbard model is already in the intermediate to strong coupling regime, for which very few methods yield reliable results. Our work showcases the potential of TPSC as a theoretical approach capable of providing reasonably accurate descriptions of the Hubbard model even at fairly strong interactions and when the frustration is present.
\end{abstract}
\maketitle

{\it Introduction.}---Quantum simulation of the 2D Fermi-Hubbard model with ultracold atomic gases has seen tremendous progress in the past decade~\cite{Gross2017,TARRUELL2018,Bohrdt2021}. These systems have long been known to offer a pristine realization of the Hubbard model through the use of optical lattices, where all the model parameters can be precisely tuned~\cite{Esslinger2010}; but it is the advent of quantum gas microscope with its unprecedented power of resolution~\cite{Kuhr2016,Schafer2020}, combined with innovative ideas of cooling~\cite{Ho2009,Bernier2009}, that has made simulating strongly correlated regimes possible. By now experiments have successfully reached temperatures as low as half of the spin exchange energy and along the way simulated many  equilibrium and non-equilibrium properties of the Hubbard model. Among other things, these include demonstration of metal-to-insulator transition~\cite{Greif2016}, probe of short- and long-range antiferromagnetic correlations~\cite{Cheuk2016,Mazurenko2017,Drewes2017,Brown2017,Xu2022}, measurement of equation of state~\cite{Cocchi2016,Hartke2020} and study of magnetic polarons~\cite{Koepsell2019,Koepsell2021,Ji2021}. In addition, various transport measurements have also been carried out in the quantum simulator, such as charge transport in the bad metallic regime~\cite{Brown2019}, spin transport of the Mott insulator~\cite{Nichols2019} as well as subdiffusion and heat transport in the high temperature regime~\cite{Sanchez2020}.  All these developments are encouraging signs that the ultimate goal of simulating the 2D Hubbard model at temperatures and dopings relevant to the physics of high Tc superconductivity may finally be within reach. 

However, the value of a quantum simulator lies not only in revealing the properties of simulated models in regimes not accessible by classical computation tools, but also in serving as an important platform for testing various theoretical ideas used to unravel these models. For the 2D Fermi-Hubbard model, an extraordinary amount of effort has been devoted to its understanding~\cite{Arovas2022} and various theoretical methods have been proposed for its quantitative description~\cite{Schafer2021,Qin2022}.  The so-called two-particle self-consistent theory (TPSC), developed by Vilk and Tremblay~\cite{Vilk1997,Tremblay2011}, is one of such methods. TPSC is a non-perturbative approach which satisfies the Mermin-Wagner theorem, the Pauli principle and various conservation and sum rules for the charge and spin. It has been applied to a wide range of problems related to the 2D Fermi-Hubbard model~\cite{Saikawa2001,Hankevych2003,Kyung2003,
Kyung2004,Tremblay2006,HANKEVYCH2006,Davoudi2006,Dare2007,Hassan2008,Bergeron2011,
Otsuki2012,Arya2015,Nishiguchi2018,Zantout2021,Martin2023,lessnich2023,gauvinndiaye2023} and was found to agree with Monte Carlo simulations on the calculation of various physical quantities in the weak-to-intermediate coupling regime~\cite{Tremblay2011}, including the double occupancy, spin and charge structure factors, and single-particle spectral functions. More importantly, it predicts the opening of pseudogap~\cite{Vilk1997} and the existence of d-wave superconductivity in the model~\cite{Kyung2003}, two very consequential results.  Although calculations based on TPSC have been directly compared to measurements of cuprate superconductors~\cite{Kyung2004,Tremblay2006,HANKEVYCH2006}, the agreement or lack thereof should be viewed along with the fact the single-band Hubbard model cannot possibly capture the real materials perfectly. For this reason, it is important to benchmark the methods proposed to solve the Hubbard model, such as the TPSC, against a quantum simulator of the model. 

In this letter, we perform such a benchmarking  for the first time by comparing the magnetic correlations calculated by TPSC to those measured in the quantum simulator of the doped and frustrated Hubbard model on an anisotropic triangular lattice~\cite{Xu2022}. 
To our surprise, we find that even for Hubbard interactions comparable to or slightly larger than the bandwidth, TPSC still gives a fairly accurate account of the magnetic correlations at almost all levels of doping and frustration. 
Our results highlight the capacity of TPSC in understanding strongly-correlated regimes of the Hubbard model that very few approaches can access.  Within quantum simulation our work also suggests a new avenue of using quantum simulators to systematically benchmark theoretical approaches on strongly correlated models.

\begin{figure}[t]
\begin{centering}
\includegraphics[width=8.4cm]{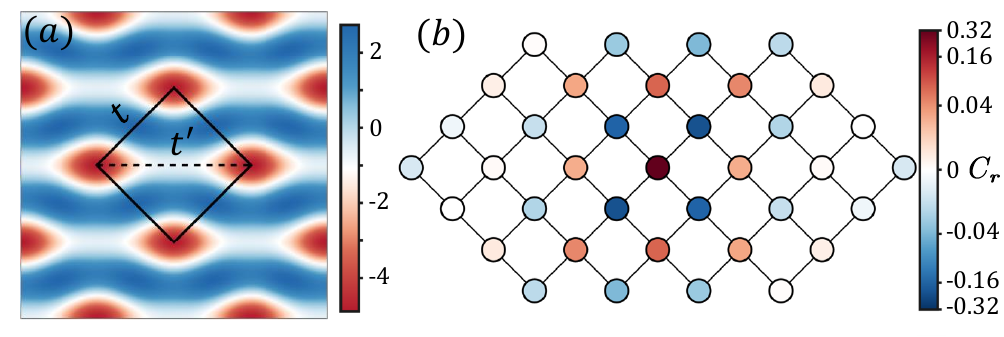}
\par\end{centering}
\caption{(a) Optical lattice potential used in Ref.~\cite{Xu2022} to realize the quantum simulator of a tunable frustrated Hubbard model. (b) A sample of the measured real-space magnetic correlation in the quantum simulator of Ref.~\cite{Xu2022}. Here $t'/t = 0.57 $, T/t = 0.34 and $U/t$ = 8.2.}
\label{fig:potential}
\end{figure}

{\it Quantum simulator of a frustrated Hubbard model.}---In the quantum simulator~\cite{Xu2022}, a two-component Fermi gas is loaded into a non-separable square optical lattice potential, formed by the interference pattern of two laser beams. When the intensities of the two beams are equal, this setup realizes the standard Hubbard model on a square lattice with nearest-neighbor hopping $t$ and on-site interaction $U$.  Through varying the relative intensity of the two beams, the horizontal diagonal hopping $t'$ can be tuned from zero to $t$ continuously, thereby realizing a tunable triangular lattice potential (see Fig.~\ref{fig:potential}(a)). The Hamiltonian of the Hubbard model on such a lattice reads
\begin{equation}
\hat H=-t\sum_{\la ij\ra\sigma}\hat c_{i\sigma}^{\dagger}\hat c_{j\sigma}-t'\sum_{\la\la ij\ra\ra,\sigma} \hat c_{i\sigma}^{\dagger}\hat c_{j\sigma} +U\sum_{i}\hat n_{i\uparrow}\hat n_{i\downarrow},\label{eq:H}
\end{equation}
where $\la\la ij\ra\ra$ denotes the next-nearest neighbor along the horizontal diagonal line with respect to the square lattice. 

In these experiments, quantum gas microscope provides an in-situ, site-resolved imaging of the atomic distribution after a spin-selective removal of one component and in so doing measure the real-space magnetic correlation function
\begin{align}
C_{\bm{r}}=4\left(\langle \hat S_{j}^{z} \hat S_{i}^{z}\rangle-\langle \hat S_{j}^{z}\rangle\langle \hat S_{i}^{z}\rangle\right ),
\end{align}
where $\br$ denotes the position of site-$j$ relative to site-$i$ and $\hat S_{i}^z = \frac{1}{2}(\hat n_{i\uparrow}-\hat n_{i\downarrow})$. A sample data of thus measured magnetic correlation from this quantum simulator is shown in Fig.~\ref{fig:potential}(b).

{\it TPSC and magnetic correlations.}---For completeness, we review the basic ideas of TPSC using the  Luttinger-Ward functional formalism~\cite{Tremblay2011}. In this formalism, the self-energy $\Sigma_{\sigma}[G_{\sigma}]$ is treated as a functional of the single-particle Green's function $G_{\sigma}$. Once a specific ansatz for the self-energy is given by some conserving scheme, the single-particle Green's function is determined by Dyson's equation. Preceding from here, however, are two independent routes to obtaining the two-particle Green's functions. One is through Heisenberg equation of motion which relates the single-particle Green's function to the two-particle one. The other is through Bethe-Salpeter equation, which expresses the susceptibilities, essentially the two-particle Green's function, in terms of the effective interactions (i.e., the irreducible vertices) $\Gamma_{\sigma\sigma'}=\delta \Sigma_{\sigma}/\delta G_{\sigma'}$. The central idea of TPSC is to enforce the consistency in the calculations of the two-particle Green's function on the choice of the variational self-energy ansatz. 

TPSC adopts a simple Hartree-like ansatz for the self-energy which assumes momentum- and frequency-independent irreducible vertices $\Gamma_{\uparrow\uparrow}$ and $\Gamma_{\uparrow\downarrow}$. In terms of the spin vertex $\Gamma_{sp}\equiv \Gamma_{\uparrow\downarrow}- \Gamma_{\uparrow\uparrow}$, the single-particle Green's function is given by
\begin{align}
G^{(1)}_\sigma(\bk,i\omega_m) = \frac{1}{i\omega_m - (\epsilon_\bk+n\Gamma_{sp}/2-\mu)}
\label{G}
\end{align}
where $\omega_m = (2m+1)\pi T$ is the fermonic Matsubara frequency at temperature $T$, 
$
\epsilon_{\bk} = -2t(\cos k_x + \cos k_y) - 2t' \cos(k_x + k_y)
$
is the single-particle dispersion, $n$ is the density and $\mu $ is the chemical potential. The equation of motion for the single-particle Green's function gives the following relation between the spin vertex $\Gamma_{sp}$ and the double occupancy $\langle \hat n_{i\uparrow}\hat n_{i\downarrow}\rangle$
\begin{equation}
\Gamma_{sp}\langle \hat n_{i\uparrow}\rangle\langle \hat n_{i\downarrow}\rangle=U{\langle \hat n_{i\uparrow}\hat n_{i\downarrow}\rangle}.\label{eq:Gammasp}
\end{equation}
The Bethe-Salpeter equation, which determines the spin susceptibility, leads to another relation between them
\begin{align}
\frac{T}{N}\sum_{\nu}\sum_{\bq}\frac{\chi^{(1)}(\bm{q},i\omega_{\nu})}{1-\frac{1}{2}\Gamma_{sp}\chi^{(1)}(\bm{q},i\omega_{\nu})}
 = n - 2{\langle \hat n_{i\uparrow}\hat n_{i\downarrow}\rangle},
 \label{eq:sumrule}
\end{align}
where $N$ is the number of lattice sites, $n$ is the density, $\omega_\nu = 2\nu \pi T$ is the bosonic Matsubara frequency and 
\begin{align}
 \chi^{(1)} \equiv\frac{T}{N}\sum_{m\bk\sigma}G^{(1)}_\sigma (\bk,i\omega_m)G^{(1)}_\sigma (\bk+\bq,i\omega_m+i\omega_\nu)
 \end{align}
is the so-called polarization bubble. For the single-particle Green's function given in Eq.~(\ref{G}), the latter takes the form of the Lindhard function $\chi^{(1)}(\bq,i\omega_\nu)=-\frac{2}{N}\sum_{\bm{k}}\frac{f(\tilde\epsilon_{\bm{k}})-f(\tilde\epsilon_{\bm{k}+\bm{q}})}{i\omega_{\nu}+\tilde\epsilon_{\bm{k}}-\tilde\epsilon_{\bm{k}+\bm{q}}}$, where $\tilde \epsilon_\bk = \epsilon_\bk + n\Gamma_{sp}/2$ and $f(\tilde \epsilon_{\bm{k}}) = 1/(e^{(\tilde\epsilon_\bk-\mu)/T} +1)$ is the Fermi-Dirac distribution.

\begin{figure}[t]
\begin{centering}
\includegraphics[width=8.7cm]{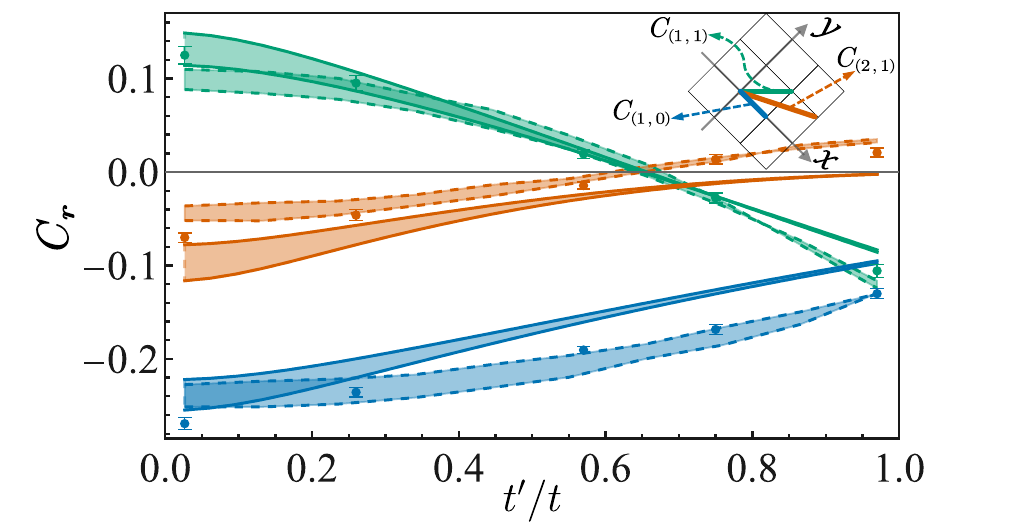}
\par\end{centering}
\caption{The real-space magnetic correlation $C_{\br}$ between the nearest-, next-nearest and next-next-nearest neighbors, denoted by $C_{(1,0)}$, $C_{(1,1)}$ and $C_{(2,1)}$ respectively.  For reference, $C_{(1,0)}=-1$ for a perfect antiferromagnetic N{\'e}el order. The dots with error bars are experimental data,  the shaded bands with solid line boundaries are TPSC calculations and those with dashed lines are DQMC simulations. Here $U/t = 9.5$, $T/t = 0.3\sim 0.35$ and $n=1$.  }
\label{fig:Cr}
\end{figure}

To obtain the relation in Eq.~(\ref{eq:sumrule}), one first notes that the spin susceptibility can be obtained from the Bethe-Salpeter equation as
\begin{align}
\chi_{sp}(\bq,i\omega_\nu) = \frac{\chi^{(1)}(\bm{q},i\omega_{\nu})}{1-\frac{1}{2}\Gamma_{sp}\chi^{(1)}(\bm{q},i\omega_{\nu})},
\end{align}
which in turn determines the spin structure factor 
\begin{align}
\tilde C_\bq  = {T}\sum_{\nu} \chi_{sp}(\bq,i\omega_\nu).
\label{STF}
\end{align}
Using the latter to calculate the real-space magnetic correlation function 
\begin{align}
C_{\br} =\frac{1}{N}\sum_{\bq} \tilde C_{\bq} e^{i \bq \cdot \br}
\label{STFF}
\end{align}
and observing that $C_{\br = 0} = n-2\langle \hat n_{i\uparrow}\hat n_{i\downarrow}\rangle$, one arrives at Eq.~(\ref{eq:sumrule}). Once $\Gamma_{sp}$ is determined self-consistently by solving Eqs.~(\ref{eq:Gammasp})-(\ref{eq:sumrule}), the spin structure factor in Eq.~(\ref{STF}) and the real-space magnetic correlation function in Eq.~(\ref{STFF})  can then be compared to experimental measurements. Lastly, we point out that the complete theory of TPSC involves additional steps to obtain a more accurate single-particle Green's function, but we omit this discussion as it is not relevant to our present purpose.

\begin{figure}[t]
\begin{centering}
\includegraphics[width=8.7cm]{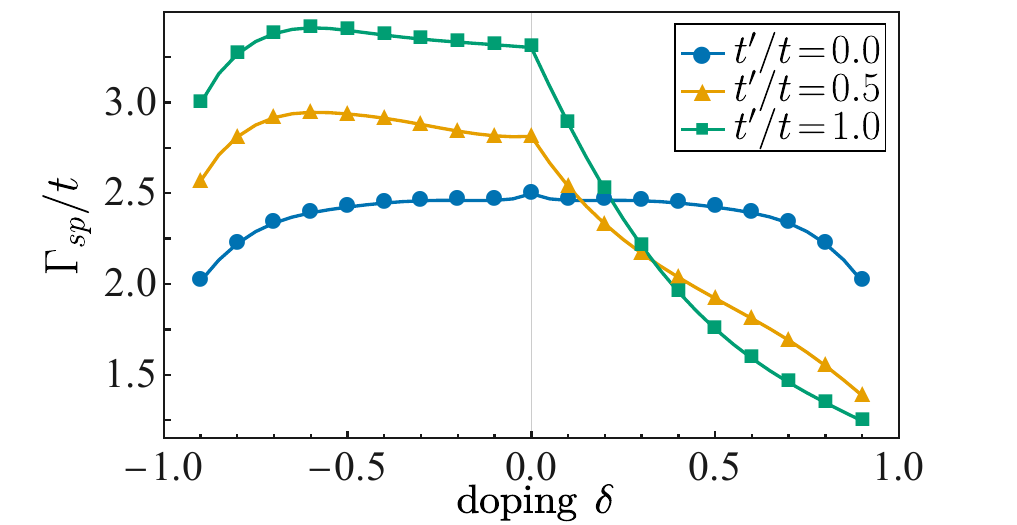}
\par\end{centering}
\caption{The spin vertex $\Gamma_{sp}$ as a function of doping at various degrees of frustration characterized by $t'/t$. Here $U/t = 9.5$, $T/t = 0.35$ and $n=1$.}
\label{Gammasp}
\end{figure}

{\it Benchmarking.}---We first compare the magnetic correlations calculated by TPSC to those measured in the quantum simulator in the case of half-filling. {To reaffirm the faithfulness of the quantum simulator we also perform determinant quantum Monte Carlo (DQMC) simulations~\cite{SM}} and include the results in the benchmarking. All TPSC calculations are done on a $256\times 256$ lattice, where sparse sampling and IR decomposition from the sparse-ir library are used to treat the temperature Green's functions~\cite{Shinaoka2017,Li2020,WALLERBERGER2023}.  As mentioned earlier, the experiments measure directly the real-space correlation function $C_{\br}$ between any two sites. Shown in Fig.~\ref{fig:Cr} are comparisons between theory and experiment on the nearest-, next-nearest and next-next-nearest neighbor correlations as a function of $t'/t$ at $U/t = 9.5$. As the final temperature of the atomic system varies slightly from one value of $t'/t$ to another, we show TPSC calculations (shaded bands in Fig.~\ref{fig:Cr}) for a range of temperatures $T/t = 0.3\sim 0.35$, consistent with that determined in the experiments. We note that these temperatures are slightly lower than the spin exchange energy $J=4t^2/U$. As we can see, the overall agreement between experiment and theory is excellent. At half-filling, the suppression of antiferromagnetic correlation by increasing frustration is clearly reflected by the reduction of the spin vertex in the TPSC calculations as shown in Fig.~\ref{Gammasp}. {However, we notice that the performance of TPSC declines slightly as frustration increases; in particular, the next-next-nearest neighbor correlation $C_{(2,1)}$ does not turn positive for large $t'/t$ as observed both in the quantum simulator and in DQMC.}

\begin{figure*}[t]
\begin{centering}
\includegraphics[width=16.6cm]{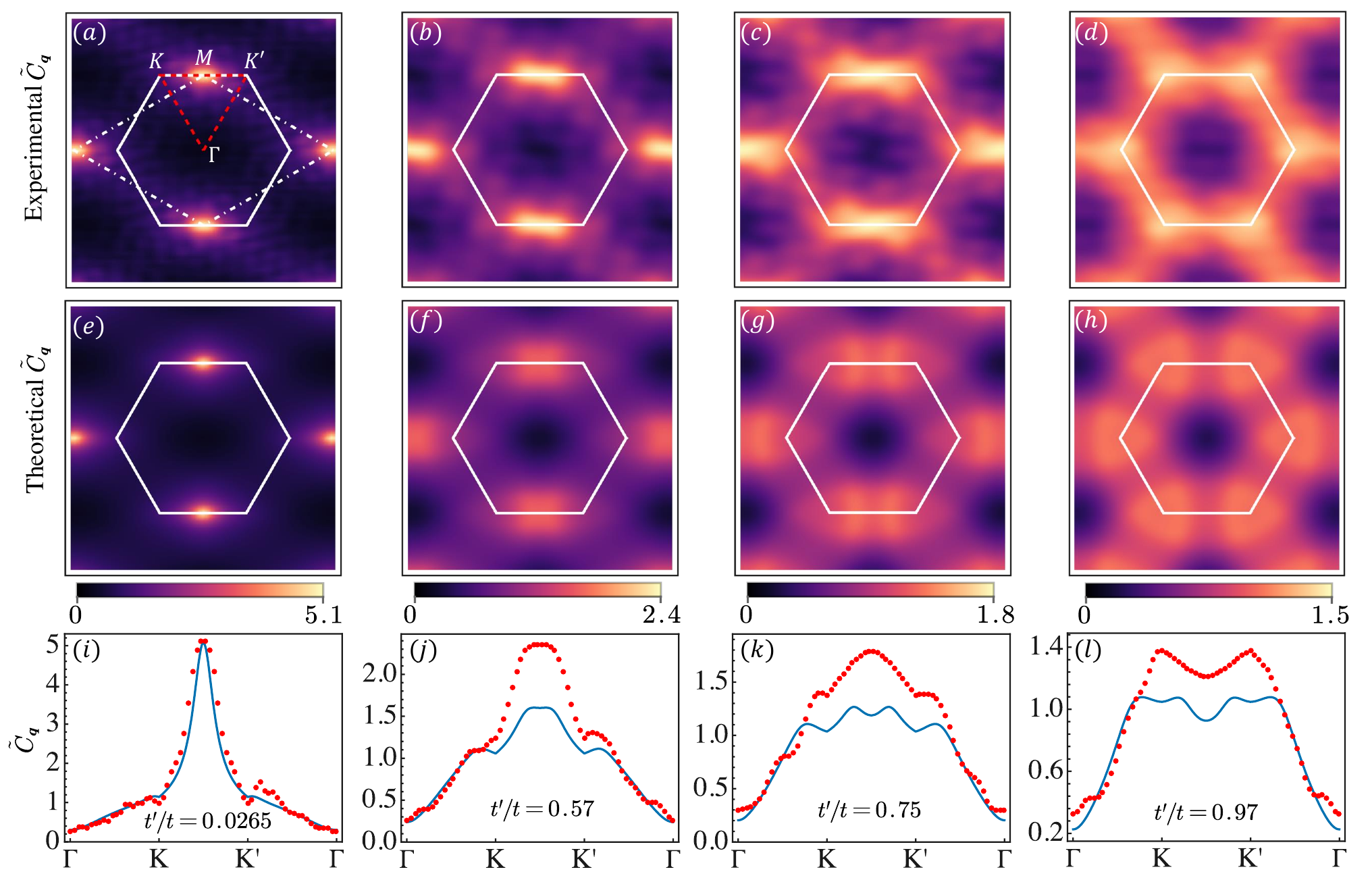}
\par\end{centering}
\caption{Spin structure factor of the Hubbard model at half-filling and in increasing degree of frustration characterized by $t'/t$. The top panel contains results from the quantum simulator, the middle panel calculations from the TPSC and the bottom panel comparisons between them. The dots in the bottom panel are experimental data and lines TPSC calculations. Here experimental parameters corresponding to different $t'/t$ are  $T/t = 0.26$ and  $U/t=9.7$ for $t'/t=0.0265$; $T/t = 0.34$ and $U/t=8.2$  for $t'/t=0.57$;   $ T/t=0.32$ and $U/t=8.2$ for $t'/t=0.75$; and    $ T/t=0.39$ and $U/t=9.2$ for $t'/t=0.97$. All the TPSC calculations in the paper are done using experimental parameters except for the case of $t'/t=0.0265$, where we find an almost perfect agreement with the experiment using a theoretical $T/t = 0.39$.}
\label{fig:STF}
\end{figure*}
\begin{figure*}[!t]
\begin{centering}
\includegraphics[width=17.2cm]{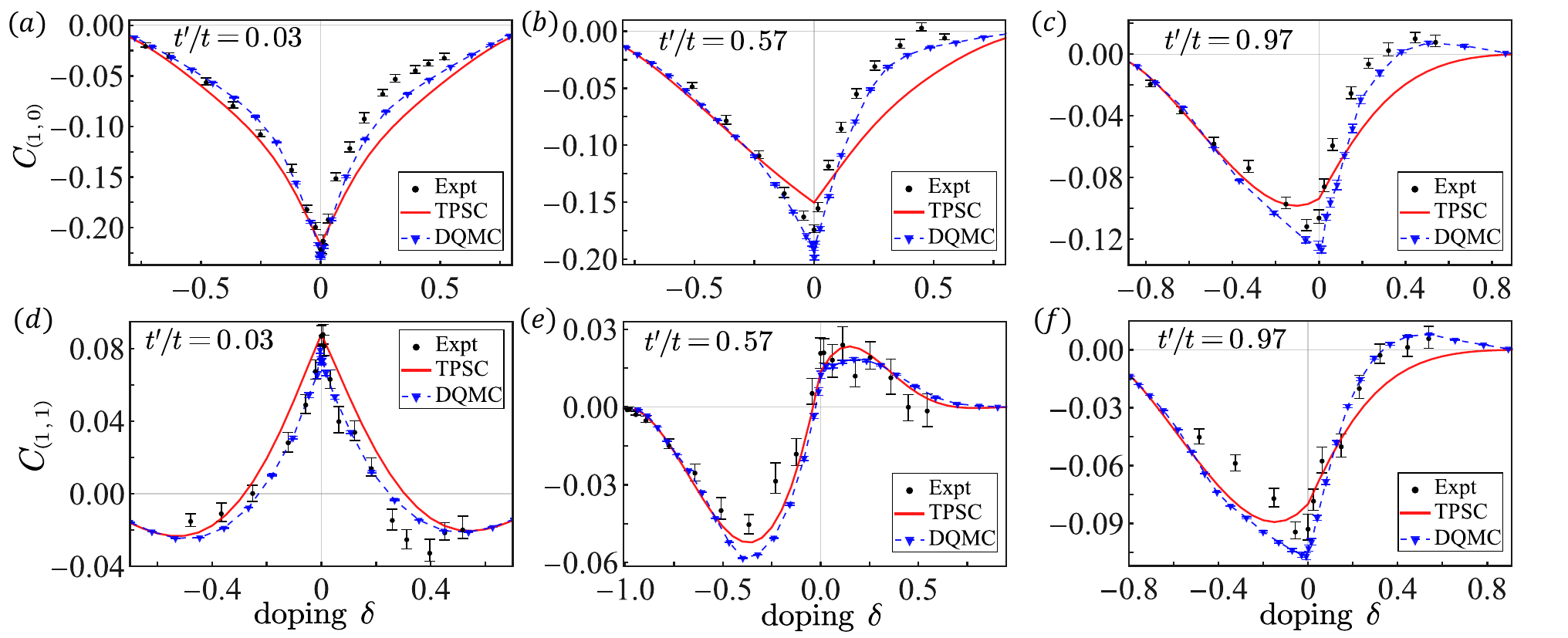}
\par\end{centering}
\caption{The nearest- and next-nearest-neighbor correlations, $C_{(1,0)}$ and  $C_{(1,1)}$, as a function of doping  and at various degrees of frustration characterized by $t'/t$. Here the  parameters used in our calculations are $U/t = 9 $, and $T/t = 0.35$ for (a)-(c) and $T/t = 0.4$ for (d)-(f). }
\label{C10doping}
\end{figure*}

 In Ref.~\cite{Xu2022}, the measured real-space magnetic correlations between all sites are then used to construct the spin structure factor $\tilde C_\bq$ by inverting  Eq.~(\ref{STFF}). There, to illustrate clearly the effects of frustration, the physical lattice in Fig.~\ref{fig:potential} is mapped to a triangular lattice  such that the square-shaped first Brillouin zone of the physical lattice now stretches into the rhombus-shaped one as illustrated in Fig.~\ref{fig:STF}(a). Plotted as a function of the crystal momentum corresponding to the triangular lattice, the experimental $\tilde C_\bq$ and the TPSC calculations are shown in Fig.~\ref{fig:STF}. Here the TPSC reproduces the full spin structure factor of the Hubbard model simulator reasonably well; in particular, the splitting of the $\tilde C_\bq$ peak at the $M$ point into two parts at the $K$ points as the frustration increases is well captured by the TPSC.  We note that TPSC does produce a smaller peak value of the spin structure factor in general compared to the experimental measurements, which could be due to the limitation of the theory. Another possible explanation is that because of the system's non-uniform density the experimental determination of the spin structure factor involves a cutoff of real-space correlations beyond a certain distance, which may result in a less accurate peak value~\footnote{Muqing Xu, private communications}. {This truncation might also account for why the  spin structure factor from the quantum simulator does not display additional oscillations as shown by TPSC (see Fig.~\ref{fig:STF}(l))~\cite{SM}.} 

We next consider the case of finite doping where the particle density is given by $n = 1+\delta$. The comparisons between the TPSC calculations and the experimental results on the nearest and next-nearest neighbor correlations as a function of doping are shown in Fig.~\ref{C10doping}; again the agreements are on the whole very good. Here an important characteristic of frustrated lattices is the lack of particle-hole symmetry in contrast to the square lattice. This can be seen from the fact that the system with particle doping $\delta>0$ can be mapped to one with hole doping $-\delta$ and described by the same Hamiltonian with $t'$ replaced by $-t'$~\cite{LiPRB2023,samajdar2023nagaoka,samajdar2023polaronic}. In fact, the magnetic correlations with particle doping are calculated  using this mapping. The observed particle-hole asymmetry in frustrated lattices is well captured by the TPSC calculations and, from the theoretical perspective, is due to the reduction of the spin vertex by particle doping in the presence of frustration (see Fig.~\ref{Gammasp}). We notice that the experimental results in Fig.~\ref{C10doping}(c) has a somewhat more pronounced particle-hole asymmetry than those of the TPSC calculations. In particular, the agreement between theory and experiment is almost perfect for the hole doping while some discrepancy exists for the particle doping. However, this discrepancy is perhaps not as large as it appears if we take into account the fact that the experimental results for the square lattice already exhibit a slight but noticeable particle-hole asymmetry (see Fig.~\ref{C10doping}(a)), due presumably to system errors. {Nevertheless, the comparison in Fig.~\ref{C10doping}(c) clearly shows that TPSC does not capture the ferromagnetic correlations observed for the isotropic triangular lattice in the vicinity of particle doping $\delta = 1/2$. This may be an indication that the assumption of constant effective interactions from TPSC is not adequate here. Since in this case a van Hove singularity exists at the Fermi level in the density of states of the non-interacting system, any momentum dependence of the effective interactions will be accentuated by the presence of this singularity and can no longer be neglected~\cite{SM}.}

{\it Conclusions.}---We have systematically benchmarked the TPSC calculations of the magnetic correlations of a doped and frustrated Hubbard model against the experimental results of a quantum simulator for $U/t = 8\sim 10$. Although TPSC is believed to be valid only for weak-to-intermediate coupling, i.e., for $U/t \lesssim 8$, the overall excellent agreement with the copious data from the quantum simulator indicates a wider range of applicability for this theory.  In the future, it would be highly desirable to test other predictions of the TPSC, such as  the opening of the pseudogap, in the next-generation of quantum simulators of the Hubbard model where the temperature can be further lowered. This can in principle be carried out by probing the single-particle spectral weight of the atomic system using the photoemission spectroscopy~\cite{Dao2007,Stewart2008,Gaebler2010}, a tool analogous to the ARPES in solid-state systems~\cite{Damascelli2003}.

\textit{Acknowledgement}. We are grateful to Dr.~Muqing Xu and Prof.~Markus Greiner for providing us the experimental data in Ref.~\cite{Xu2022}. We also want to thank Dr. Muqing Xu and Dr. Lev Kendrick for stimulating discussions. This work is supported by National Key R$\&$D Program of China (Grant No. 2022YFA1404103), NSFC (Grant No.~11974161) and Shenzhen Science and Technology Program (Grant No.~KQTD20200820113010023).

\bibliography{BibSI}

\end{document}


\title{Supplemental Material for ``Magnetic correlations of a doped and frustrated Hubbard model: benchmarking the two-particle self-consistent theory against a quantum simulator"}
\author{Guan-Hua Huang}
\affiliation{Hefei National Laboratory, Hefei 230088, China}
\author{Zhigang Wu}
\email{atticuspku@gmail.com}
\affiliation{Quantum Science Center of Guangdong-Hong Kong-Macao Greater Bay Area (Guangdong), Shenzhen 508045, China}
\date{\today }

\maketitle
This supplemental material includes the following three sections: (I) Determinant quantum Monte Carlo simulation; (II) Spin structure factor and (III) Ferromagnetic correlations. 
 
\section{Determinant Quantum Monte Carlo simulation}
Determinant quantum Monte Carlo (DQMC) method, also known the auxiliary field quantum Monte Carlo method, is a finite-temperature algorithm developed for the simulation of models of interacting electrons on a lattice, such as the Hubbard model~\cite{test}.
For the purpose of reaffirming the faithfulness of the quantum simulator in Ref.~\cite{Xu2023}, we perform DQMC simulations for all the system parameters probed by the quantum simulator using the SmoQyDQMC.jl package~\cite{10.21468/SciPostPhysCodeb.29}. 
The simulations are performed on an $8\times 8$ lattice with the imaginary time Trotter step size $ d\tau = \beta/N_{\tau}\approx 0.02/t$.  As DQMC simulations encounter the negative sign problem at low temperatures in the frustrated Hubbard model, sufficient statistical averaging is needed to reduce the computation error. Specifically, for temperatures $T/t = 0.3$, we average over $40$ independent runs with different seeds to obtain reliable statistics where each run is generated with $5000$ warm-up passes and $30000$ measurement passes; for temperatures T/t = 0.35, we average over $20$ runs where each run is generated with $5000$ warm-up passes and $20000$ measurement passes; for temperatures $T/t = 0.4$, we average over $20$ runs where each run is generated with $5000$ warm-up passes and $15000$ measurement passes.

\section{Spin structure factor}
\begin{figure}[thb]
	\centering
	\includegraphics[width = 17cm]{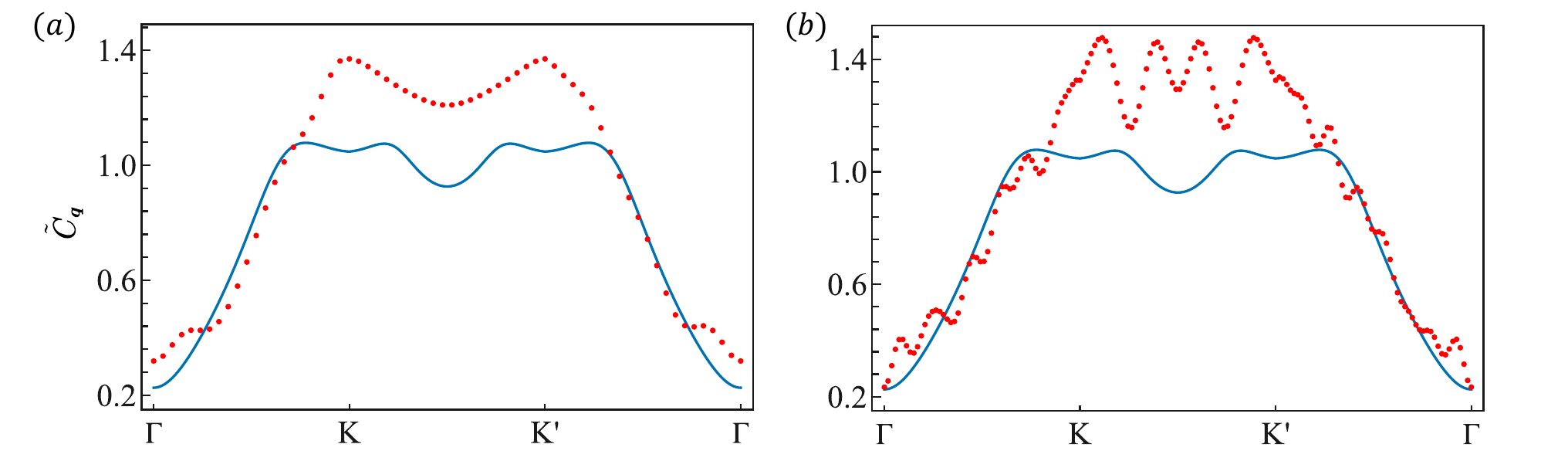}
	\caption{(a) Comparison between the spin structure factor from TPSC (solid line) and that from the quantum simulator (dots) with truncated Fourier sum.  (b) Comparison between the spin structure factor from TPSC (solid line) and that from the quantum simulator (dots) with complete Fourier sum.  } 
	\label{ssftruc}
\end{figure}

Experimentally,  the spin structure factor is obtained by Fourier transforming the real-space spin correlation function $C_{\br}$, i.e.,  
\begin{align}
  \tilde C_{\bq} = \sum_{\br}^{|\br| \le d_{max}} C_{\br} e^{i\bq\cdot\br},
\end{align}
where the sum is truncated at a maximum distance $d_{max}$. Although the experiments measure real-space spin correlations up to a distance of $d = 11$ (in units of lattice spacing), the experimental spin structure factor in Fig.~\ref{ssftruc}(a) is obtained by taking $d_{max} = 5$ (see {\bf Methods} section of Ref.~\cite{Xu2023}). The reason is that experimentalists regard the measured real-space correlations beyond $d_{max} = 5$ as noises~\footnote{Muqing Xu, private communications} and thus discarded them. However, if we use the complete experimental data to compute the spin structure factor instead, we find a result shown in Fig.~\ref{ssftruc}(b). Interestingly, this spin structure factor exhibits oscillations just like the TPSC results, although the peaks are more pronounced than those in TPSC. In our view, the existence of the additional oscillations from TPSC reflects the fact that the spin structure factor obtained using this approach contains contributions from long-distance spin correlations. 

\section{Ferromagnetic correlations}
We have seen in the main text that TPSC fails to capture the ferromagnetic correlation observed in the quantum simulator for the triangular lattice in the vicinity of particle doping $\delta = 1/2$.
\begin{figure}[h]
	\centering
	\includegraphics[width=17cm]{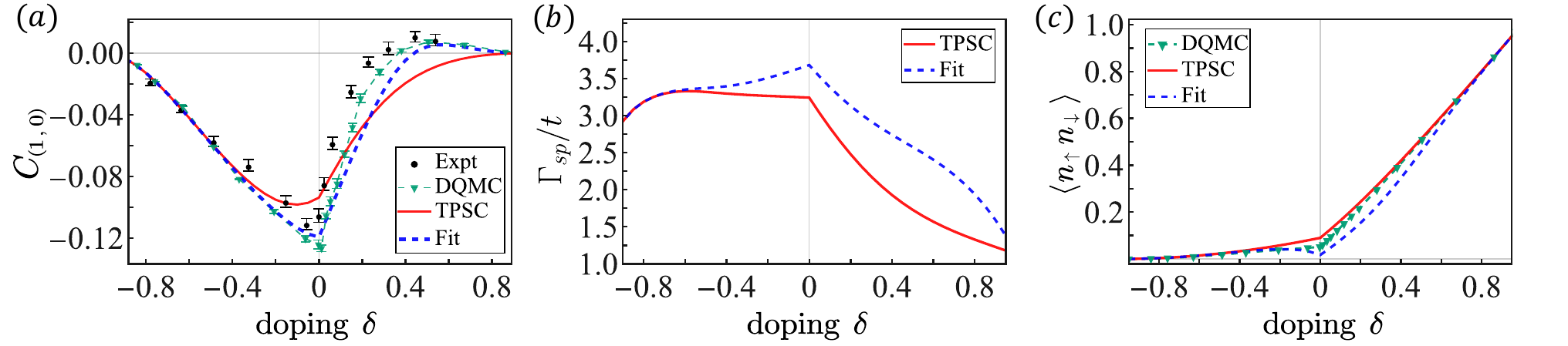}
	\caption{(a) The nearest-neighbor correlation $C_{(1,0)}$ from TPSC, TPSC with a fitted $\Gamma_{sp}$ shown in (b), the quantum simulator and DQMC. Here $t'/t = 0.97$, $U/t = 9 $, and $T/t = 0.35$. (b) $\Gamma_{sp}$ obtained from TPSC and $\Gamma_{sp}$ that yields the dashed (blue) curve in (a). (c) The double occupancy obtained from TPSC, DQMC and the TPSC with a fitted $\Gamma_{sp}$. }
	\label{fitting}
\end{figure}
 As shown in Fig.~\ref{fitting}(a) here, the nearest-neighbor correlation $C_{(1,0)}$ from the quantum simulator and DQMC become positive at particle doping $\delta = 1/2$ for $t'/t = 0.97$ while the TPSC result does not. However, if we treat $\Gamma_{sp}$ as a free fitting parameter in Eq.~(7) of the main text, we can obtain ferromagnetic correlations  at particle doping $\delta = 1/2$ by choosing appropriate  $\Gamma_{sp}$ values. For example, if we choose $\Gamma_{sp}$ as indicated by the blue dashed line in Fig.~\ref{fitting}(b), the resulting $C_{(1,0)}$ can actually capture the ferromagnetic correlation as shown in Fig~\ref{fitting}(a). But this new $\Gamma_{sp}$ leads to a double occupancy that deviates further away from the original TPSC and DQMC values as shown by Fig~\ref{fitting}(c). This clearly shows the inadequacy of assuming a momentum and frequency independent $\Gamma_{sp}$ in this case. 

\begin{figure}[htb]
\centering
\includegraphics[width=17cm]{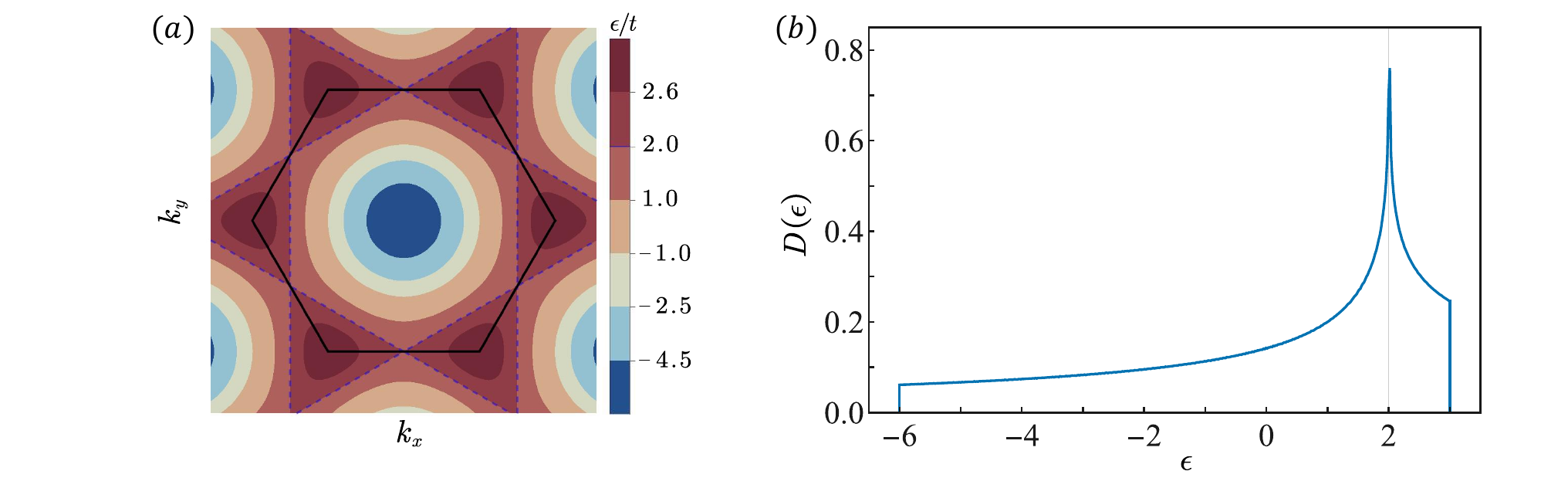}
\caption{(a) The first Brillouin zone of non-interacting band structure of the isotropic triangular lattice, where the dashed line shows the Fermi surface corresponding to particle doping at $\delta = 1/2$. (b) The density of states of the non-interacting band structure of the isotropic triangular lattice, where the van Hove singularity occurs at the Fermi level $\mu =2t $. }
\label{dos}
\end{figure}
It is intriguing that the case of  isotropic triangular lattice around particle doping of $\delta = 1/2$  is special in this sense. We postulate that it might be related to the fact that there is a van Hove singularity at Fermi level in the non-interacting density of states, which is shown in Fig.~\ref{dos}.
To see this we first note that if we retain the momentum and frequency dependence of the spin vertex $\Gamma_{sp}(\bq, i\omega_\nu)$, the spin susceptibility is given by
\begin{align}
\chi_{sp}(\bq,i\omega_\nu) = \frac{\chi^{(1)}(\bq,i\omega_{\nu})}{1-\frac{1}{2}\Gamma_{sp}(\bq, i\omega_\nu)\chi^{(1)}(\bq,i\omega_{\nu})},
\end{align}
from which the real-space magnetic correlation function can be determined as
\begin{align}
C_{\br} &=\frac{T}{N}\sum_{\bq,\nu} \chi_{sp}(\bq,i\omega_\nu)e^{i \bq \cdot \br} \nonumber\\ &= \frac{T}{N}\sum_{\bq,\nu}\frac{\chi^{(1)}(\bq,i\omega_{\nu})}{1-\frac{1}{2}\Gamma_{sp}(\bq, i\omega_\nu)\chi^{(1)}(\bq,i\omega_{\nu})}e^{i \bq \cdot \br} \nonumber\\
& = T\sum_\nu \int d\epsilon D(\epsilon) \frac{\chi^{(1)}(\bq,i\omega_{\nu})}{1-\frac{1}{2}\Gamma_{sp}(\bq, i\omega_\nu)\chi^{(1)}(\bq,i\omega_{\nu})}e^{i \bq \cdot \br},
\label{STFF}
\end{align}
where $D(\epsilon)$ is the density of states of the non-interacting system. Since $\Gamma_{sp}({\bq,i\omega_\nu})$ depends on $\epsilon $ through $\bq$, 
any momentum-dependence of $\Gamma_{sp}(\bq,i\omega_\nu)$ will be magnified by the presence of the singularity and can no longer be neglected.   It would be an important goal for our future work to generalize TPSC further by including the momentum and frequency-dependence of the spin vertex in a way which does not violate the two-particle self-consistency. 


\bibliography{BibSISM.bib}